\documentclass[pra,aps,twocolumn]{revtex4-1}
\usepackage{amsmath}
\usepackage{amssymb}
\usepackage{graphicx}
\DeclareGraphicsExtensions{.eps}

\newcommand{\mP}{{\mathcal P}}
\newcommand{\mT}{{\mathcal T}}

\newcommand{\beqa}{\begin{eqnarray}}
\newcommand{\eeqa}{\end{eqnarray}}

\begin{document}
\title{Tunable waveguide lattices with non-uniform parity-symmetric tunneling} 
\author{Yogesh N. Joglekar}
\author{Clinton Thompson}
\author{Gautam Vemuri}
\affiliation{Department of Physics, 
Indiana University Purdue University Indianapolis (IUPUI), 
Indianapolis, Indiana 46202, USA}
\date{\today}
\begin{abstract}
We investigate the single-particle time evolution and two-particle quantum correlations in a one-dimensional $N$-site lattice with a site-dependent nearest neighbor tunneling function $t_\alpha(k)=t_0[k(N-k)]^{\alpha/2}$. Since the bandwidth and the energy levels spacings for such a lattice both depend upon $\alpha$, we show that the observable properties of a wavepacket, such as its spread and the relative phases of its constitutents, vary dramatically as $\alpha$ is varied from positive to negative values.  We also find that the quantum correlations are exquisitely sensitive to the form of the tunneling function. Our results suggest that arrays of waveguides with position-dependent evanascent couplings will show rich dynamics with no counterpart in present-day, traditional systems. 
\end{abstract}
\maketitle

\noindent{\it Introduction:} Idealized lattice models have been popular in physics due to their analytical and numerical tractability~\cite{wen}, the absence of divergences associated with the ultraviolet cutoff~\cite{kogut1,kogut2}, the availability of exact solutions~\cite{onsager}, and the ability to capture counter-intuitive physical phenomena including the bound states in repulsive potentials~\cite{winkler}. Over the years, these models have been successful in describing a diverse array of physical systems with bosons, fermions, and quantum spins, with short- or long-ranged interactions, such as electronic materials, optical lattices~\cite{bloch,zoller}, and, most recently, evanescently coupled optical waveguides~\cite{review}. The prototypical lattice models have a constant nearest neighbor tunneling amplitude, and the effects of ubiquitous disorder and imperfections are taken into account via random impurity potentials and small, random variations in the tunneling amplitude. They are sufficient to capture important physical phenomena such as Anderson localization~\cite{anderson}. 

In recent years, coupled optical waveguides have become a paradigm for the realization of an ideal one-dimensional lattice model with tunable tunneling and on-site potential, as well as non-Hermitian parity- and time-reversal ($\mP\mT$) -symmetric potentials~\cite{bendix,longhi1}. They have been used demonstrate several phenomena from condensed matter physics and quantum optics, such as Bloch oscillations~\cite{bloch2}, Dirac zitterbewegung~\cite{longhi3}, Talbot effect~\cite{talbot}, and quantum random walks~\cite{qrw}. Anderson localization due to random on-site potential, introduced by a randomly varying refractive index, has been experimentally observed in waveguides with a constant nearest-neighbor tunneling~\cite{berg1}. Two-particle Anderson localization, quantum statistics effects, and quantum and classical correlations have been theoretically explored in such waveguides as well~\cite{berg2}. The spontaneous $\mP\mT$-symmetry breaking has been observed in two coupled waveguides with $\mP\mT$-symmetric complex index of refraction~\cite{ruter}. Most of these cases, with the notable exception of Refs.~\cite{gf,longhi2}, have primarily focused on on-site disorder effects in a one-dimensional lattice with roughly constant nearest-neighbor hopping that, in the continuum limit, translate into disorder effects on particle with a finite mass and a quadratic dispersion; in particular, properties of itinerant quantum particles in a lattice with position-dependent tunneling amplitude have not been extensively explored.

In this paper, we show that a one-dimensional lattice with position-dependent tunneling function $t_\alpha(k)=t_0[k(N-k)]^{\alpha/2}=t_\alpha(N-k)$ has a rich dynamics with no counterpart in the traditional lattice. This model is motivated by the robust $\mP\mT$-symmetric phase of its non-Hermitian counterpart~\cite{ya}. The results that we report here on single-particle propagation and localization, and two-particle correlations in such lattices can be investigated in an array of coupled optical waveguides.  Physically, one would need to engineer the waveguides such that the coupling between adjacent waveguides has the form mentioned above.  Globally, we find that the parameter $\alpha$ has a significant effect on the evolution of the input wavepacket, and that tuning $\alpha$ allows one to tailor certain aspects of the wavepacket evolution.  

Our salient results are as follows: i) the spread of a wavepacket, after propagating a certain distance along the waveguides, monotonically depends upon $\alpha$; in particular, when $\alpha\lesssim -1$, the wavepacket spread is negligible for physical propagation distances. ii) when $\alpha=1$, the phase-information in the initial state of the particle is accessible only within windows around certain propagation distances, and the size of these windows can be controlled by the location of the input waveguide; when $\alpha\neq 1$, this phase information is, in principle, always accessible. iii) for two quantum particles injected into adjacent waveguides, the two-particle correlation function is exquisitely sensitive to $\alpha$ and the location of the input waveguide. 

As we discuss below, these results are not substantially affected by a "weak" disorder. They show that coupled optical waveguides with specifically engineered tunneling functions may provide novel, heretofore unexplored, realizations of lattice models with tunable energy levels, densities of states~\cite{ya}, and two-particle correlations. 

\noindent{\it Tight-binding Model:} We consider an array of $N$ waveguides described by the Hamiltonian with open boundary conditions, 
\begin{equation}
\label{eq:tb}
H_\alpha=-\sum_{i=1}^{N-1}t_{\alpha}(i) \left(a^{\dagger}_{i+1}a_i + a^{\dagger}_i a_{i+1}\right)+\sum_{i=1}^Nv_i a^{\dagger}_ia_i,
\end{equation}
where $a^{\dagger}_k$ is the creation operator for a particle at site $k$, $t_\alpha(k)$ is tunneling amplitude between sites $k$ and $k+1$, and $v_k$ represents the potential, determined by the local index of refraction,  at site $k$. The tunneling amplitude $t(k)$ is determined by the evanescent coupling between waveguides $k$ and $k+1$, and can be tuned by varying the width of the barrier between the two waveguides~\cite{gf}. A Hamiltonian eigenfunction $|\psi^n\rangle=\sum_k\psi^n_k a^{\dagger}_k |0\rangle$ with energy $E^{n}$ satisfies the difference equation
\begin{equation}
\label{eq:recursion}
t_\alpha(k-1)\psi^{n}_{k-1} + t_\alpha(k)\psi^{n}_{k+1}=-E^{n}_\alpha \psi^{n}_k,
\end{equation}
where we have considered a constant index of refraction $n_R$, which results in a constant shift in the energy eigenvalues. The eigenvalue spectrum for Eq.~\ref{eq:recursion} is symmetric about zero~\cite{yogesh}. Hence, the  bandwidth of the spectrum, defined as the difference between the maximum and minimum eigenvalues, is $\Delta_\alpha= E_\mathrm{max}-E_\mathrm{min}=2E_\mathrm{max}$. Note that when $\alpha>0$, the tunneling function $t_\alpha(k)$ is maximum at the center of the waveguide array whereas when $\alpha<0$, it is maximum at the ends. As a result, when $N\gg1$ the bandwidth $\Delta_\alpha(N)$ of the Hamiltonian $H_\alpha$ increases monotonically with $\alpha$. It is natural to use the inverse-bandwidth as the characteristic time, $T_\alpha=2\hbar/\Delta_\alpha$, and $L_\alpha= cT_\alpha/n_R$ as the characteristic distance along the waveguide where $\hbar=h/(2\pi)$ is the scaled Planck constant and $c/n_R$ is the speed of light in a waveguide. Note that since $T_\alpha$ and $L_\alpha$ are monotonically decreasing functions of $\alpha$, a waveguide array with a fixed physical length will correspond to "short-time" scenario when $\alpha<0$ and "long-time" scenario when $\alpha>0$. 

\begin{figure}[t!]
\begin{center}
\includegraphics[angle=0,width=11cm]{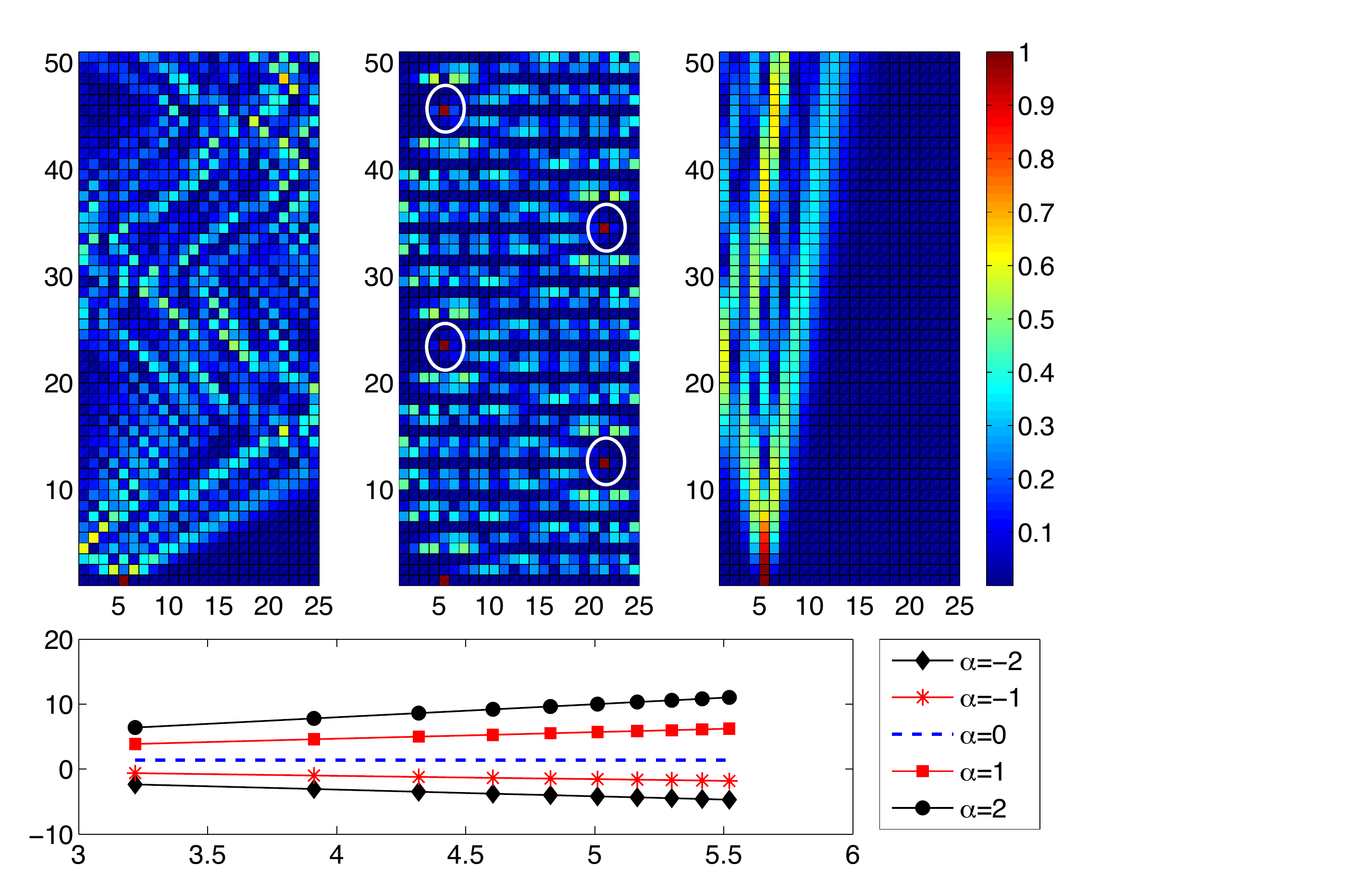}
\caption{(color online) Time evolution in an array with a fixed length $L/(\hbar c/n_Rt_0)=50$ shown along the vertical axis. Top three panels show the probability-amplitude time-evolution plots of a single photon injected in waveguide $m_0=5$ in an array of $N=25$ waveguides with $\alpha=0$ (left panel), $\alpha=+1$ (center panel) and $\alpha=-1$ (right panel). The vertical axis in each panel represents the distance along the waveguide. When $\alpha=0$ the wavepacket, initially localized at $m_0=5$, spreads as it travels along the waveguide. For $\alpha=1$, because the energy levels are equidistant, the wavepacket is periodically localized at mirror symmetric positions $(N+1-m_0)=21$ and $m_0=5$ (white circles). When $\alpha=-1$, the wavepacket spread is noticably smaller over the same length of the waveguide. The bottom panel shows the dimensionless bandwidth $\Delta_\alpha(N)/t_0$ (vertical) vs. $N$ (horizontal) on a logarithmic scale for $25\leq N\leq 250$. We see that when $\alpha\geq 1$, $\Delta_\alpha(N)\propto N^\alpha$, whereas for $\alpha\leq-1$, $\Delta_\alpha(N)\propto N^{-\alpha/2}$. Therefore, a sample with a given physical length represents "short-time" evolution when $\alpha<0$ and "long-time" evolution when $\alpha>0$. }
\label{fig:lvsalpha}
\end{center}
\vspace{-5mm}
\end{figure}
Figure~\ref{fig:lvsalpha} shows the time-evolution of a wavepacket that is initially localized in waveguide $m_0=5$ in an array of $N=25$ waveguides. The vertical axis denotes distance along the waveguide {\it for a fixed physical length of the waveguide} $L/(\hbar c/nt_0)=50$. The three vertical panels correspond to $\alpha=0$ (left), $\alpha=1$ (center) and $\alpha=-1$ (right). When $\alpha=0$, the traditional model, the wavepacket broadens as it travels down the length of the waveguide array. When $\alpha=1$, the energy levels are given by $E_n=\pm t_0(N-1), \pm t_0(N-3),\ldots$; the level spacing is constant and the bandwidth is $\Delta_{\alpha=1}(N)=2(N-1)t_0$~\cite{ya,longhi2}. Therefore we obtain perfect reconstruction of the wavepacket, shown by white circles, at mirror-symmetric positions $(N+1-m_0)=21$ and $m_0=5$. It should be noted that the wavepacket first reconstructs at the mirror symmetric waveguide, i.e. 21st site, and then alternates between sites 5 and 21.  For $\alpha=-1$ (right panel), {\it the wavepacket spread over the same distance along the waveguide is significantly smaller}, consistent with the smaller bandwidth of the Hamiltonian.

\begin{figure}[t!]
\begin{center}
\includegraphics[angle=0,width=11cm]{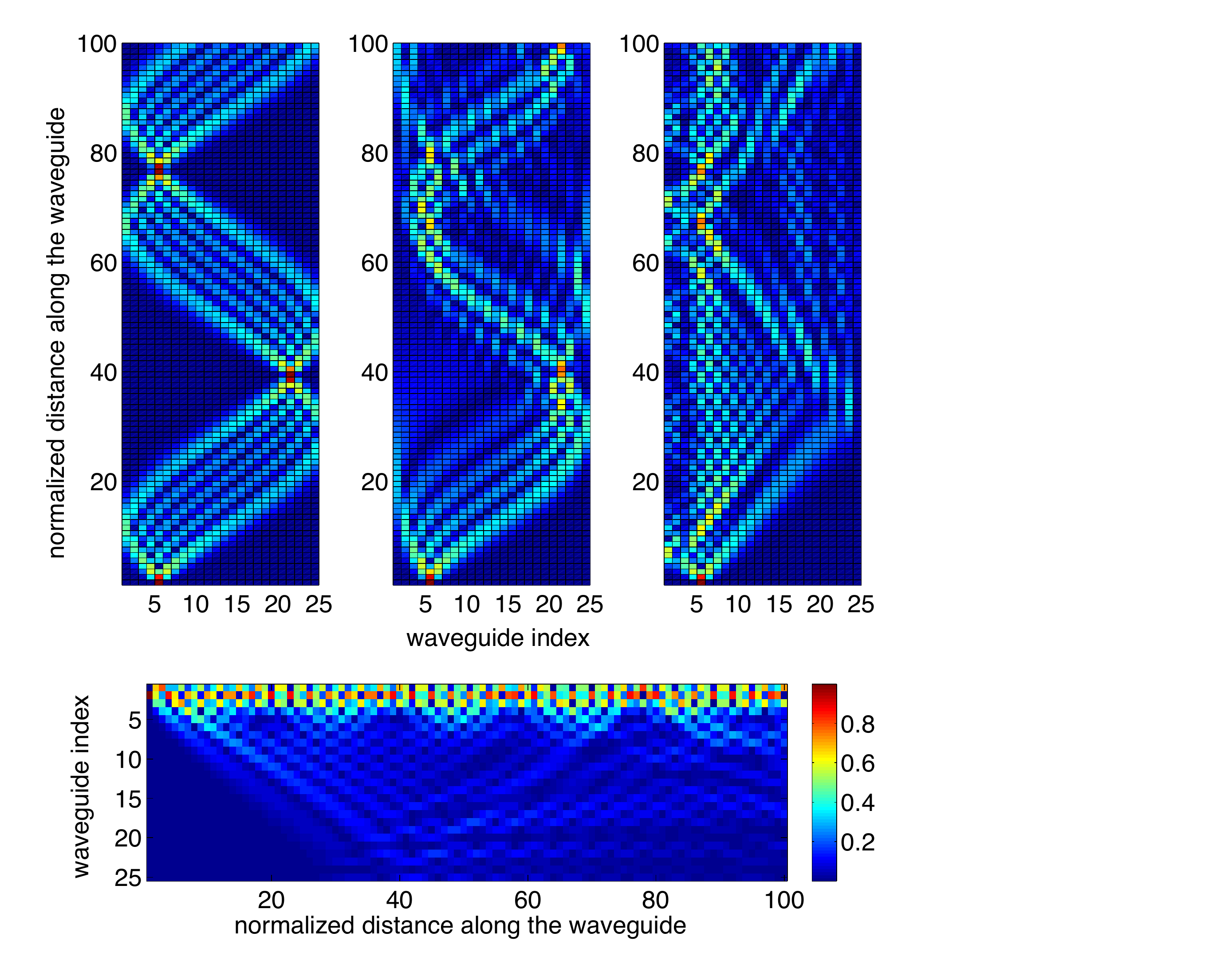}
\caption{(color online) Top panels show the time evolution of a photon injected at $m_0=5$ in an array of $N=25$ waveguides with $\alpha=1$ (left), $\alpha=2$ (center), and $\alpha=-1$ (right). The wavepacket spread is similar over normalized length-scales. The bottom panel shows, for $\alpha=-1$, the time evolution of a photon injected near the edge, $m_0=2$. The strong localization of the photon near the edge is due to the presence of localized edge eigenstates that occur when $\alpha<0$.}
\label{fig:loc}
\end{center}
\vspace{-5mm}
\end{figure}

To explore the intrinsic $\alpha$-dependence of the time-evolution, in the rest of the paper, we consider waveguides with the {\it same normalized length} $L/L_\alpha=100$; physically, this will correspond to waveguides with different $\alpha$-dependent lengths. Figure~\ref{fig:loc} shows the time-evolution of a wavepacket initially at $m_0=5$ for $\alpha=1$ (left panel), $\alpha=2$ (center panel) and $\alpha=-1$ (right panel) in an array of $N=25$ coupled waveguides. The vertical-axis shows distance (time) in the units of $L_\alpha (T_\alpha)$. Apart from the perfect reconstruction at mirror-symmetric points that occurs when $\alpha=1$, we see that, in contrast to the behavior in Fig.~\ref{fig:lvsalpha} the spread of the wavepacket is qualitatively similar for all $\alpha$ over the normalized length-scales (or time-scales). The bottom panel shows, for $\alpha=-1$, the time-evolution of a single photon injected near the edge, $m_0=2$; the horizontal axis shows the normalized distance (time). In this case, the photon remains at the edge due to localized edge eigenstates that are generically present when $\alpha<0$~\cite{ya}. 

\begin{figure}[h!]
\begin{center}
\hspace{5mm}
\includegraphics[angle=0,width=12cm]{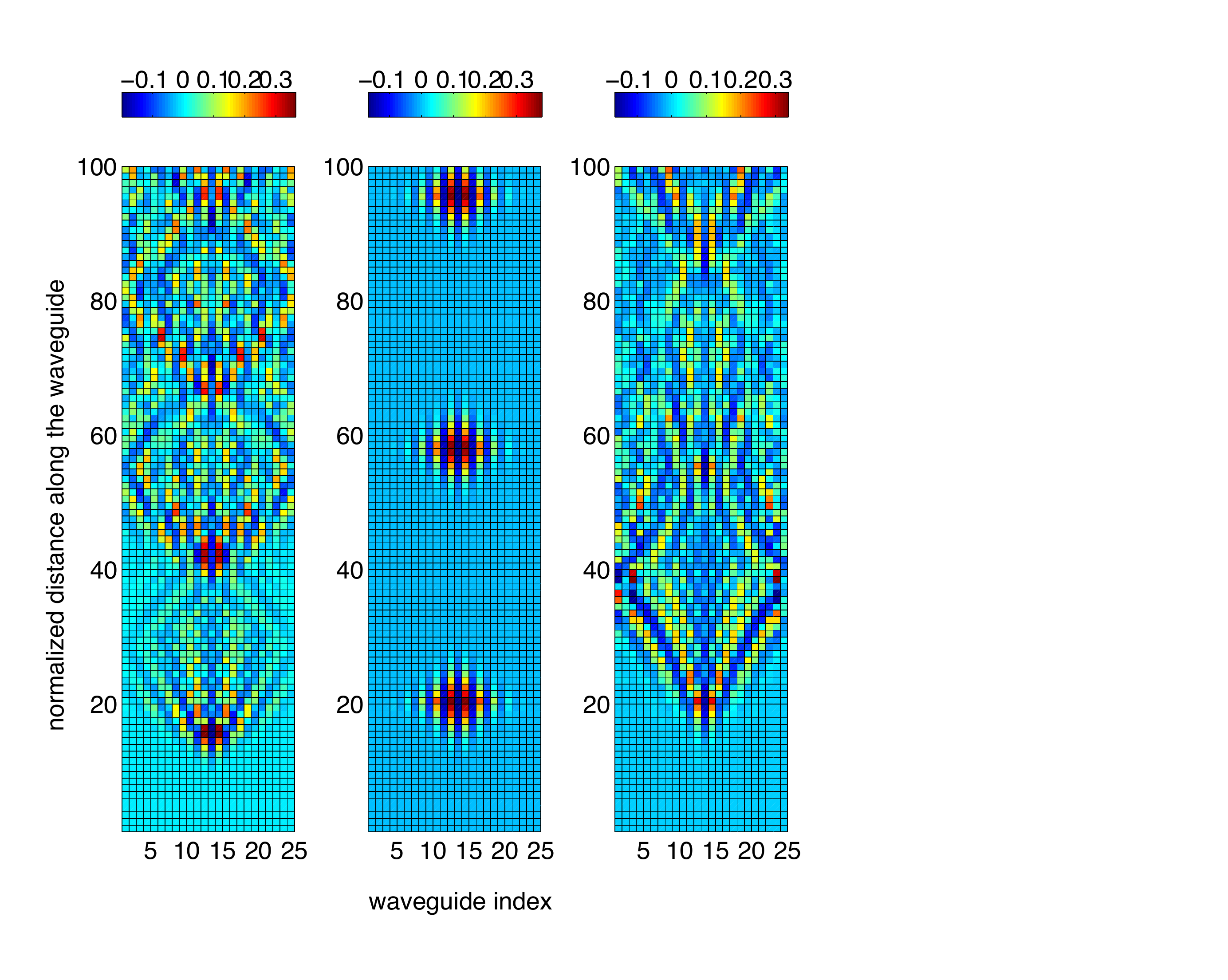}
\caption{(color online) Maximal amplitude difference ${\mathcal A}(t,k)$ for an array of $N=25$ waveguides with initial state $|\psi_\phi\rangle=(|1\rangle + e^{i\phi} |N\rangle)/\sqrt{2}$. In all cases, ${\mathcal A}=0$ at short distances $l/L_\alpha\lesssim 15$ where the partial waves from initial waveguides do not interact with each other. When $\alpha=0$ (left panel) and $\alpha=-1$ (right panel) the phase information persists, as is expected for a clean system. When $\alpha=1$ (center panel), however, the phase information is accessible only in certain (diamond shaped) windows.}
\label{fig:phase}
\end{center}
\vspace{-5mm}
\end{figure}

Lastly, we consider a wavepacket that is initally localized in mirror-symmetric waveguides, $|\psi_\phi\rangle=(|m_0\rangle + e^{i\phi} |N+1-m_0\rangle)/\sqrt{2}$. We obtain the time-evolution of the $\phi$-dependent probability amplitude $A(\phi,t,k)=|\langle k|\psi_\phi(t)\rangle|$ and use the maximal difference ${\mathcal A}(t,k)=A(0,t,k)-A(\pi/2,t,k)$ as the indicator of the phase information. Note that since the initial state is localized in two spatially separated regions, information about the phase $\phi$ will become visible in ${\mathcal A}(t,k)$ only after a time when the partial waves from the two mirror-symmetric sites interfere with each other.  Figure~\ref{fig:phase} shows ${\mathcal A}(t,k)$ for an array of $N=25$ waveguides with $m_0=1$, where the vertical axis represents normalized distance (time) along the waveguide. When $\alpha=0$ (left panel) and $\alpha=-1$ (right panel) the phase information, indicated by a nonzero value of ${\mathcal A}$, is visible at all times, as is expected for a clean system. When $\alpha=1$ (center panel), however, the phase information is available only in a restricted window in the $(t,k)$ space. The size of this window increases with $m_0$. Thus, when $\alpha=1$, the {\it information about the initial relative phase remains inaccessible over a large fraction} of the parameter space. 

\noindent{\it Two-particle correlations:} We now explore the effects of the tunneling function $t_\alpha(k)$ on the two-particle (number) correlation function defined by $\Gamma^\alpha_{mn}(t)=\langle a^{\dagger}_m(t) a^{\dagger}_n(t)a_n(t)a_m(t)\rangle$. This function encodes the Hanbury-Brown-Twiss quantum correlations in coincidence detections in waveguides $m$ and $n$~\cite{hbt}. For an initial state where the two particles are localized at sites $(m_0,n_0)$, the correlation function becomes 
\begin{equation}
\label{eq:gamma}
\Gamma^\alpha_{mn}(t)=|G_{mm_0}(t) G_{nn_0}(t)\pm G_{mn_0}(t) G_{nm_0}(t)|^2.
\end{equation}
where $G_{pq}(t)=[\exp(-iH_\alpha t/\hbar)]_{pq}$ is the time-evolution operator and $\pm$ signs correspond to bosons and fermions respectively. When $\alpha=0$, the traditional model, properties of this correlation function and its dependence on the initial state have been extensively investigated~\cite{berg2}. Since $\Gamma^\alpha_{mn}(t)$ is determined by the time-evolution operator, it follows that the bosonic and fermions correlations will be qualitatively different when $\alpha\neq 0$. In particular, when $\alpha=1$, the constant energy level spacing implies that $\Gamma^{\alpha=1}_{mn}(t)$ is periodic in time or, equivalently, in the distance along the waveguide; since the maximum spread of a wavepacket initially confined at position $1\leq m_0\leq N/2$ is approximately $2m_0$, it follows that the spatial extent and shape of the correlation function in the $(m,n)$ plane can be controlled by appropriate initial conditions. 

\begin{figure}[t!]
\begin{center}
\includegraphics[angle=0,width=9cm]{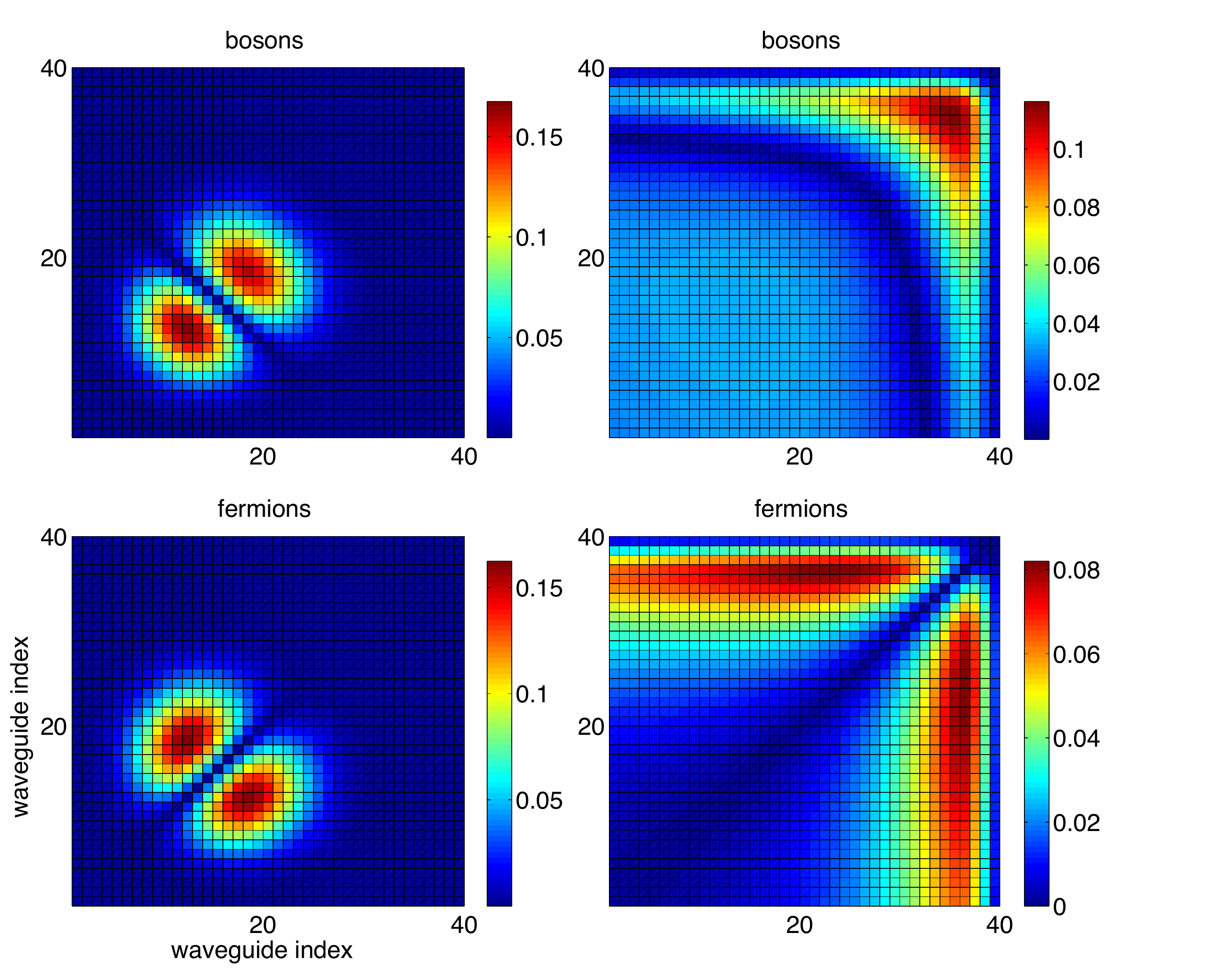}
\caption{(color online) Left panels show the correlation function $\Gamma^\alpha_{mn}(t)$ for an array of $N=40$ waveguides at $t/T_\alpha=25$ and $\alpha=1$.  The initial state of the system has two particles in the first two waveguides. The correlation function remains localized, and develops only two symmetric peaks with a single nodal line. Right panels correspond to $\alpha=2$ and $t/T_\alpha=55$ with the same initial conditions. We see that the bosonic correlation function (top) is localized near the second edge, whereas its fermionic counterpart (bottom) is localized in one direction and extended in the other direction.}
\label{fig:corr}
\end{center}
\vspace{-5mm}
\end{figure}
Figure~\ref{fig:corr} shows $\Gamma^\alpha_{mn}$ for an array with $N=40$ waveguides and $(m_0,n_0)=(1,2)$. The left panels show the results for $\alpha=1$ for bosons (top) and fermions (bottom) at time $t/T_\alpha=25$. In contrast to the $\alpha=0$ case~\cite{berg2}, the correlation function is strongly localized at all times, and has only two peaks with a single nodal line separating them. The right panels correspond to $\alpha=2$ and $t/T_\alpha=55$. At this time, the {\it bosonic} correlation function (top) is {\it localized near the second edge}, with a nearby parabolic nodal region. On the other hand, the {\it fermionic} correlation function (bottom) is sharply {\it localized in one direction and extended in the other}, with a broad nodal region around the diagonal. (We recall, from the central panel in Fig.~\ref{fig:loc}, that when $\alpha=2$, a wavepacket starting near the edge localizes substantially near the other edge when $t/T_\alpha\sim 50$.) These results show that the quantum statistics lead to nontrivial correlations for $\alpha=2$ case that are dramatically different from the $\alpha=0$ case~\cite{berg2} or the $\alpha=1$ case. 

\noindent{\it Discussion:} In this paper, we have shown that modifying the tunneling function in a tight-binding Hamiltonian, which can be realized by an array of coupled waveguides, produces a wide range of wavepacket evolutions that are not seen in traditional models. The tunneling function $t_\alpha(k)$ affects the wavepacket properties through the bandwidth $\Delta_\alpha$ and energy level spacings, both of which are dependent on $\alpha$. 

For waveguides with a fixed length, we have shown that there are qualitative differences in the wavepacket time-evolution depending on whether $\alpha$ is positive or negative. For example, when $\alpha=1$, the equidistant energy levels lead to periodic behaviors such as wavepacket reconstruction~\cite{longhi2}; when $\alpha<0$, a wavepacket near the edge remains localized due to edge eigenstates. In addition, we have shown that when $\alpha=1$ the phase-information about an initial state remains inaccessible over a large region of the parameter space, whereas when $\alpha\neq 1$, it is accessible. 

We have shown that the tunneling function modifies quantum correlations in a non-trivial manner. For example, when $\alpha=1$, the size and the shape of bosonic and fermionic correlations can be tuned by the choice of initial waveguides; the periodicity of these correlations follows from the equidistant energy spectrum. For the same initial conditions, when $\alpha=2$, we find that the correlations, including the no-coincidence region for bosons and fermions, are dramatically different.

These result are applicable for a clean, disorder-free system. For a finite lattice, a weak disorder $v_d/\Delta_\alpha\ll 1$, will localize a wavepacket to its initial waveguide~\cite{berg1} after a sufficiently long time $T_l\gg T_\alpha$, or distance along the waveguide. The disorder and propagation-distance thresholds, as well as the effect of a weak nonlinearity, however, depend upon $\alpha$~\cite{clint}. Our results, thus, remain valid at times $T_\alpha\lesssim t\ll T_l$.  

C.T. was supported by a GAANN award from the US Department of Education to G.V.

\end{document}